\documentclass[twocolumn, pra, aps, amsfonts, amsmath, amssymb]{revtex4-1}
\usepackage{graphicx}
\usepackage{footmisc}
\usepackage{bbm}
\usepackage{subcaption}
\usepackage{physics}
\usepackage{physics}
\graphicspath{ {./images/} }
\usepackage[utf8]{inputenc}
\usepackage[export]{adjustbox}
\usepackage{enumerate}
\usepackage{float}
\everymath{\displaystyle}
\usepackage{comment}

\usepackage[colorlinks=true,bookmarks=false,citecolor=blue,urlcolor=blue]{hyperref} 

\newcommand{\orcid}[1]{\href{https://orcid.org/#1}{\includegraphics[width=10pt]{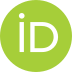}}}

\usepackage[customcolors]{hf-tikz}

\tikzset{style green/.style={
    set fill color=green!20,
    set border color=green!20,
  },
  style blue/.style={
    set fill color=blue!20,
    set border color=blue!20,
  },
   style red/.style={
    set fill color=red!20,
    set border color=red!20,
  },
  hor/.style={
    above left offset={-0.15,0.2},
    below right offset={0.175,-0.1},
    #1
  },
  ver/.style={
    above left offset={-0.075,0.2},
    below right offset={0.075,-0.1},
    #1
  }
}

\usepackage{multirow}

\begin{document}

\title{Robust entanglement measure for mixed quantum states }

\author{Dharmaraj Ramachandran\orcid{0000-0002-3068-1586}}
\email{p20200040@goa.bits-pilani.ac.in}
\author{{Aditya Dubey}\orcid{0009-0008-9444-8506}}\email{f20200627@goa.bits-pilani.ac.in}
\author{Subrahmanyam S. G. Mantha\orcid{0009-0001-0589-5689}}\email{f20200502@goa.bits-pilani.ac.in}
\author{Radhika Vathsan\orcid{0000-0001-5892-9275}}
\email{radhika@goa.bits-pilani.ac.in}

\affiliation{Physics Department, BITS Pilani K K Birla Goa Campus}

\begin{abstract}
 We introduce an entanglement measure, the Modified Bloch Norm (\( MBN \)), for finite-dimensional bipartite mixed states, based on the improved Bloch matrix criteria~\cite{shen2016improved}. \( MBN \) is demonstrated to be effective in analyzing the dynamics of bound entanglement—a valuable resource for quantum protocols where free entanglement may not be available.  
Through examples, we illustrate the applications of \( MBN \) in accurately estimating the Entanglement Sudden Death (ESD) time and detecting behavior such as the freezing of entanglement. Additionally, we show that the error rate for entanglement measured using state estimation from a limited number of measurement copies is significantly lower when using \( MBN \) compared to negativity. This demonstrates the robustness of \( MBN \) under practical constraints.

\end{abstract}

\maketitle

\section{Introduction}

Entanglement is recognized as one of the most significant features of quantum systems and serves as a valuable resource contributing to the quantum advantage in various information-theoretic protocols \cite{horo_review}. This importance underscores the need to quantify the entanglement of a quantum state. Although numerous entanglement measures exist for pure states, it has been conclusively proven that determining the entanglement of an arbitrary mixed quantum state is an NP-hard problem \cite{np-hard-sep, np-hard-sep1}. Given the computational intractability of a complete solution, substantial research has focused on making entanglement quantification feasible under certain assumptions or constraints. 

In this work, we introduce an entanglement measure that addresses two practical challenges in the entanglement quantification of mixed states: measuring bound entanglement, and estimating entanglement from measurements on a finite number of copies of the state.

\subsection{Dynamics of bound entanglement}
 A finite-dimensional separable bipartite quantum state on a Hilbert space \( H_A \otimes H_B \) is represented by a density matrix that is a linear combination of separable density matrices \cite{werner_sep}:
\begin{equation}
    \rho_{\small AB} = \sum_i p_i \rho_{A,i} \otimes \rho_{B,i}.
    \label{ent}
\end{equation}
A bipartite quantum state that cannot be expressed in the form given in Eq.~\eqref{ent} is entangled.

Entanglement in a quantum state is considered distillable  or free if using  \( n \) copies of the state, one can obtain   \( k \leq n \) copies of maximally entangled states, using Local Operations and Classical Communications (LOCC)\cite{horodecki2001distillation}. Entangled states that possess non-distillable entanglement are known as bound entangled states \cite{bound_ent_98}, which are also useful across a range of applications \cite{useful_bound_2019, bounduse1, bounduse2, bounduse3, bounduse4, bounduse5, bounduse6, bounduse7}. For instance, in 2006, Masanes \textit{et al.} showed that all entangled bipartite states, whether bound or distillable, can be used in teleportation protocols \cite{bound_ent_usedful_2006}.

While the dynamics of distillable entanglement under various noisy channels have been studied extensively \cite{dyn1, dyn2, dyn3, dyn4}, very little is known about the dynamics of bound entanglement \cite{bound_dyn1, bound_dyn2}. In this work, we propose an entanglement measure suitable for studying the dynamics of both free and bound entangled states.

\subsection{Entanglement quantification using finite copies of quantum state}
In practice, an unknown quantum state is estimated from the expectation values of observables measured on a number of copies of the state. Error in estimation is inherent to the finite sample size taken for measurement.  The number of observables needed to characterize an \( N \) level quantum system scales as \( N^2 - 1 \), leading to an exponential increase in the sample size requirement. Consequently, the error in estimation is expected to increase. Therefore, experimenters are constrained to quantify properties of quantum states, such as entanglement, using inherently imprecise expectation values~\cite{expt_challenges}.

The simplest technique to estimate an \( N \)-dimensional quantum state is Linear Inversion (LI) \cite{LI1} using the standard representation of a state in the basis of identity and the generators of \(SU(N)\), \(\sigma_i\):
\begin{equation}
    \rho_{\text{est}} = \frac{1}{N}\left[ \mathbbm{1}_N + x_i \sigma_i\right].
    \label{herm}
\end{equation}
Here \( x_i \)'s are estimated expectation values of the \( \sigma_i \)'s, which  are traceless, Hermitian, orthogonal  \(N\times N\) operators, the natural generalizations of the Pauli operators to $N$ dimensions~\cite{blochvector_2003}.
Though computationally less demanding due to  its simplicity, LI has a major limitation: \( \rho_{\text{est}} \) could be negative, violating a fundamental condition of a density operator. 

Other standard approaches to estimate density operators rely on methods such as Maximum Likelihood Estimation (MLE)\cite{MLE1,MLE2,MLE3} and Bayesian Inference (BI)\cite{BI1,BI2}. However, both MLE and BI are computationally intensive for arbitrary quantum states. Alternative methods, such as randomized measurement, seek to quantify entanglement without requiring full state tomography \cite{randomised2,randomised1}. 
There has been promising success in quantum state estimation using machine learning \cite{ml1,ml2}, but a generalized model to faithfully estimate finite-dimensional quantum states remains a challenging task. The entanglement measure introduced in this work exhibits low error rates when based on inaccurate estimations of quantum states.

\section{Existing Entanglement measures}

There exist several entanglement measures for pure states, such as entanglement entropy \cite{ent_entr} and geometric measures based on Bures distance \cite{bures_dist}, which are straightforward to compute and satisfy the properties of an entanglement monotone. However, their extensions to mixed states are based on convex roof construction, which is computationally demanding.

Concurrence, an entanglement measure derived from entanglement of formation, has a closed-form expression for two-qubit systems \cite{concurrence}. For multipartite pure states, concurrence is generalized as tangle \cite{conc_tangle}, which quantifies only the residual entanglement in the system \cite{residual_ent, residual_ent2}. Beyond convex-roof construction and some other numerical approximation techniques \cite{mixed_state_conc, mixed_conc2, mixed_conc3, mixed_conc4}, there is no closed-form expression for calculating concurrence in arbitrary-dimensional mixed states. 

The Positive Partial Transpose (PPT) criterion is a well-known separability criterion that asserts all density matrices with a negative partial transpose are entangled \cite{ppt1, ppt2}. However, for systems of dimensions higher than \(2 \otimes 3\), the PPT criterion is only necessary but not sufficient. This limitation leads to the existence of PPT entangled states, also known as bound entangled states, as first reported by Horodecki \textit{et al.} \cite{ppt-ent}.

A well known measure constructed based on PPT criterion is negativity which is defined as \cite{negativity}, 
\begin{equation*}
    \mathcal{N}(\rho) := \sum_{a_i<0} a_i 
\end{equation*}
where $a_i$'s are the eigenvalues of the partial transposed density matrix. 

Although negativity is a widely used entanglement measure, it cannot detect bound entangled states  due to its basis in the PPT criterion. Several criteria, such as the Realignment Criterion \cite{realignment}, the Computable Cross Norm Criterion (CCNR) \cite{ccnr1, ccnr2}, and the Improved Bloch Matrix (IBM) criterion, which forms a family of entanglement detection criteria, have been proposed to detect bound entanglement. Among these, the Improved Bloch Matrix criterion stands out due to its strong relevance to experimentally measurable quantities and its flexibility. In this work, we adopt the Improved Bloch Matrix criterion to construct an entanglement measure capable of addressing this gap.

\section{Separability based on Bloch matrix picture}

Consider a bipartite composite  system  Hilbert space $\mathcal{H}_A\otimes \mathcal{H}_B$, where $\dim[\mathcal{H}_A] =N, \dim[\mathcal{H}_B] =M$. In terms of the generators $\lambda_i$  of $SU(N)$ and  $\tilde{\lambda}_j$ of $SU(M)$, we have
\begin{multline}
    \rho_{AB} = \frac{1}{NM} \left( \mathbbm{1}_N \otimes \mathbbm{1}_M + \sum_{i = 1}^{N^{2}-1} r_i\lambda_i \otimes \mathbbm{1}_{M} + \right.\\ 
   \left . \sum_{j = 1}^{M^2 - 1} s_j \mathbbm{1}_N \otimes \tilde{\lambda}_j + \sum_{i = 1}^{N^{2}-1} \sum_{j = 1}^{M^{2}-1} t_{ij}\lambda_i \otimes \tilde{\lambda}_j \right),
    \label{rho_herm}
\end{multline}
where $\lambda$ and $\tilde{\lambda}$ are traceless Hermitian orthogonal matrices\cite{Bloch-rep1,Bloch-rep2,vb_criteria}, commonly taken to be the   
generalized Gell-Mann matrices\cite{blochvector_2003}. 
The coefficients $ \{r_1 , r_2 ...r_{M^2 - 1}\}$ and $ \{s_1 , s_2 ...s_{N^2 - 1}\}$ represent the local Bloch vectors $\mathbf{r}^t$ 
 and $\mathbf{s}^t$ of A and B subsystems respectively, while $T = [t_{ij}]$ is the correlation matrix between the two subsystems. 
The local density operators are represented in terms of the Bloch vectors:
\begin{eqnarray*}
\Tr_B(\rho_{AB}) = \rho_A = \frac{1}{N}\left( \mathbbm{1}_N + \sum_ir_i \lambda_i\right ),\\
\Tr_A(\rho_{AB}) = \rho_B = \frac{1}{M}\left( \mathbbm{1}_M + \sum_i s_j \tilde{\lambda}_i\right ).
\end{eqnarray*}

The Bloch matrix representation of the  state $\rho_{AB}$ is then written as \cite{vb_criteria}
\begin{equation}
    B_{\rho_{AB}} = 
    \begin{bmatrix}
        1 & \mathbf{s}^t \\
        \mathbf{r} & T
    \end{bmatrix}.
    \label{BM}
\end{equation}
 This representation facilitates two well known separability criteria: the Correlation Matrix (CM) criterion \cite{vb_criteria} and Generalized Correlation Matrix (GCM) criterion\cite{lb-criteria}.  Later, Shen \textit{et al} gave improved separability criteria based on a modification of the Bloch matrix. We refer to this as the Improved Bloch Matrix (IBM) criteria,   which prove to be more discriminant than the CM and GCM criteria \cite{shen2016improved}.   
Their modified Bloch matrix is constructed  by prepending $m$ 
 scaled copies of the first row  and column of the Bloch matrix: 
\begin{equation}
S_{m,a,b}(\rho) = 
    \begin{bmatrix}
    a b E_{m \times m} & b w_{m}(\mathbf{s})^{t} \\
    a w_{m}(\mathbf{r}) & T
    \end{bmatrix}
    \label{MBM}
\end{equation}
where $E_{m \times m}$ is the $m \times m$ matrix of ones, 
$a$ and $b$ are positive real scale factors. $w_m(\mathbf{x})$ for an $n$-dimensional column vector $\mathbf{x}$ is   
\begin{equation*}
    w_m(\mathbf{x}) = \underbrace{\begin{bmatrix}
        \begin{bmatrix}
        x_1 \\
        x_2 \\
        . \\
        . \\
        x_n
        \end{bmatrix}& \begin{bmatrix}
        x_1 \\
        x_2 \\
        . \\
        . \\
        x_n
        \end{bmatrix} & \begin{matrix}
        . \\
        . \\
        . \\
        . \\
        .
        \end{matrix}  & \begin{matrix}
        . \\
        . \\
        . \\
        . \\
        .
        \end{matrix}  & \begin{bmatrix}
        x_1 \\
        x_2 \\
        . \\
        . \\
        x_n
        \end{bmatrix} 
            \end{bmatrix}}_{\text{m columns}}.
\end{equation*}

The IBM criteria for separability  of $\rho_{AB}$ states that \cite{shen2016improved} 
\begin{equation}
    |S_{m,a, b}(\rho)|_{\text{tr}} \leq \frac{1}{2} \sqrt{(2ma^2 + N^2 - N)(2mb^2 + M^2 - M)},
    \label{criteria}
\end{equation}
where $|\; |_{tr}$ denotes trace norm of the matrix.

The IBM criteria are necessary conditions for separability  of mixed bipartite quantum states for all $a,b \in [0 , \infty)$ and $m \in \mathbb{Z}^{+}$.

However the authors  \cite{shen2016improved} have shown that discriminance of IBM always improves when m increases by fixing $a \text{ and } b$ as $a= \sqrt{{2}/{M(M-1)}}, \;b = \sqrt{{2}/{N(N-1)}}$. Thus, for all examples in this manuscript, we have fixed $a$ and $b$ to the above mentioned values and $m$ to be $4$, which was sufficient to detect entanglement in the chosen states.

\noindent  It is worth noting that Eq-\eqref{criteria} reduces to the CM criterion at $a = b = 0$ and to GCM criterion  at $m = a = b = 1$.

\textbf{Pure states:} 
Modified Bloch matrix $S_{m , a ,b}^{s}$ of a separable pure bipartite state $\rho_{s}$ can always be expressed in terms of Bloch vectors of the subsystems as
\begin{equation*}
    S_{m , a ,b}^{s} = \begin{bmatrix}
        bE_{m \times 1}\\
        r
    \end{bmatrix}
    \begin{bmatrix}
        a E_{\small 1 \times m} & s^{t}
    \end{bmatrix}
\end{equation*}
By properties of matrix norms, the trace norm of $S^{s}_{m,a,b}$ is 
\begin{equation*}
    |S^{s}_{m,a,b}|_{tr} = \frac{1}{2} \sqrt{(2ma^2 + N^2 - N)(2mb^2 + M^2 - M)}.
\end{equation*}
Thus IBM criteria become both necessary and sufficient conditions for separability for all finite dimensional pure bipartite quantum states~\cite{shen2016improved}.

\section{Bipartite Measure}
We  quantify the violation  of  the IBM criterion exhibited by a bipartite quantum state $\rho$  by the quantity
\begin{equation}
    V_{\rho} = |S_{m, a,b}|_{\text{tr}} - c,
\end{equation}
where  $c = \frac{1}{2} \sqrt{(2ma^2 + N^2 - N)(2mb^2 + M^2 - M)}$, the right hand side of the inequality Eq.\eqref{criteria}. 
We propose a bipartite measure of entanglement, the Modified Bloch matrix Norm ($MBN$) of a bipartite quantum state on  $H^N \otimes H^M$ as 
\begin{eqnarray}
    MBN(\rho) &=& \max(0 , V_{\rho})/V(\rho_m), 
    \label{mbn}
\end{eqnarray}
where the  normalisation constant  $V(\rho_m)$ of the maximally entangled state $\rho_m$ of the given bipartition. This is a family of measures depending on the parameters $a,b$ and $m$.
\subsection*{Properties of \(MBN\)}
We now verify that $MBN$ satisfies all the requisite properties of a valid entanglement measure.

\noindent \textbf{Discriminance:} 
    An entanglement measure must be zero for a separable state. This property is known as weak discriminance:
\begin{equation*}
    E(\rho) = 0 \quad \forall \quad \rho \in \Omega_s,
\end{equation*}
where \( \Omega_s \) denotes the set of separable states. The measure is strongly discriminant if the converse is also true: if \( E(\rho) \neq 0 \) for all entangled states.

The IBM criterion Eq.\eqref{criteria} is not violated by separable states. Therefore the derived $MBN$ is inherently weakly discriminant. 

On the other hand, $IBM$ is violated by all entangled pure states but not all entangled states
as counterexamples can show. Thus $MBN$ is strongly discriminant for only pure states and weakly discriminant in general. 

 While $MBN$ can detect various bound entangled states it can neither detect all bound entangled nor freely entangled states.
Now the separability problem itself is NP-hard \cite{np-hard-sep}, and proving a criterion to be strongly discriminant is therefore also hard. In fact,  formulating a completely discriminant and computable entanglement measure for higher-dimensional mixed states is expected to be extremely challenging, if not infeasible.

\noindent \textbf{Convexity:}
Let $\rho$  be a convex combination  of two density matrices $\rho_1$ and $\rho_2$, $\rho = p\rho_1 + (1-p) \rho_2 $.  A positive function $E(\rho)$ is convex if 
\begin{equation}
    E(\rho) \leq p E(\rho_1) + (1-p) E(\rho_2).
    \label{convexity}
\end{equation}
In terms of the modified  Bloch matrices $S({\rho_1})$ and $S({\rho_2})$,  for any value of $a,b$ and $m$, the modified  Bloch matrix of  the convex combination $\rho$ is 
\begin{eqnarray*}
S({\rho}) = p  S({ \rho_1}) +  (1-p) S({\rho_2}).
\end{eqnarray*}
This is true due to the linearity of the Bloch matrix representation.
The  matrix norm is a convex function: 
\begin{equation}
 |S({\rho})|  \leq p |S({\rho_1})| + (1-p) |S({\rho_2})|.
 \label{conve}
 \end{equation}
 Therefore $MBN$ is convex.

\noindent \textbf{Invariant under Local Unitary operations:}
Let us consider a bipartite density matrix $\rho_{AB}$. From Eq: \eqref{rho_herm} we know that $\rho_{AB}$ can always be expressed in terms of generators $\lambda_i$ and $\tilde{\lambda_j}$  of $SU(N)$ and $SU(M)$ as
\begin{multline}
    \rho_{AB} = \frac{1}{NM} \left( \mathbbm{1}_N \otimes \mathbbm{1}_M + \sum_{i = 1}^{N^{2}-1} r_i\lambda_i \otimes \mathbbm{1}_{M} + \right.\\ 
   \left . \sum_{j = 1}^{M^2 - 1} s_j \mathbbm{1}_N \otimes \tilde{\lambda}_j + \sum_{i = 1}^{N^{2}-1} \sum_{j = 1}^{M^{2}-1} t_{ij}\lambda_i \otimes \tilde{\lambda}_j \right),
   \label{eq:rho}
\end{multline}
When $\rho_{AB}$ undergoes a Local Unitary(LU) transformation by $U = U_A \otimes U_B$ then the LU transformed density matrix $\rho^{\prime}_{AB}$ is expressed as 
\begin{multline}
\rho^{\prime}_{AB} = \frac{1}{NM} \left( \mathbbm{1}_N \otimes \mathbbm{1}_M + \sum_{i = 1}^{N^{2}-1} r_i\lambda^{\prime}_i \otimes \mathbbm{1}_{M} + \right.\\ 
   \left . \sum_{j = 1}^{M^2 - 1} s_j \mathbbm{1}_N \otimes \tilde{\lambda}^{\prime}_j + \sum_{i = 1}^{N^{2}-1} \sum_{j = 1}^{M^{2}-1} t_{ij}\lambda^{\prime}_j \otimes \tilde{\lambda}^{\prime}_j \right),
   \label{eq:rho_prime}
\end{multline}
where 
$
\rho^{\prime}_{AB} = U\rho_{AB}U^{\dagger} , \;\;
\lambda^{\prime}_{i} = U_{A}\lambda_iU_A^{\dagger}, \;\;
\tilde{\lambda}^{\prime}_{j} = U_{B}\lambda_iU_B^{\dagger}
$. By definition $\lambda_i^{\prime}$ and $\tilde{\lambda}_i^{\prime}$ are generators of $SU(N)$ and $SU(M)$ respectively. 

From Eq: \eqref{eq:rho} and \eqref{eq:rho_prime} it is straightforward to see that the modified Bloch Matrix of $\rho$ and $\rho^{\prime}$ are exactly the same except for the change in set of generators. 
Thus $MBN(\rho)$ is invariant under LU transformations. 


\noindent \textbf{Entanglement monotone:}
Entanglement monotones are positive functions that do not increase under probabilistic Local Operations and Classical Communications (LOCC). It is considered to be the most important property for an entanglement measure~\cite{horo_review}.

\noindent W. D\"ur \textit{et. al} in \cite{D_r_2000_3qubit} have discussed about constructing entanglement monotones based on Stoachastic Local Operations and Classical Communications (SLOCC) invariants. Later, C. Eltschka \textit{et al} extended this discussion and proved that a positive homogeneous function of pure bipartite states invariant under determinant-1 SLOCC operations defines an  entanglement monotone if and only if
the homogeneous degree is not larger than 4~\cite{Polydeg_4}.

In order to prove that $MBN$ is an entanglement monotone, we initially highlight that the trace norm of the modified Bloch matrix $S_{m,a,b}$ is invariant under determinant-1 SLOCC operations.

\noindent Jeng \textit{et al.} \cite{slocc_inv} have shown that the eigen values of the Bloch matrix is invariant under determinant-1 SLOCC operations. Below we summarize major results of \cite{slocc_inv}. 

When an SLOCC symmetry $g = A_1 \otimes A_2 \in G = SL(d_1 )\otimes SL(d_2 )$ acts on the state
space $H = H_1 \otimes H_2$, this action  $\rho 	\rightarrow g\rho g^{\dagger} = (A_1 \otimes A_2 )\rho(A^{\dagger}_1\otimes A^{\dagger}_2)$ 
can be represented by a tensor $C_k$ such that 
\begin{equation}
    A_k\lambda_i^{(k)}A_k^{\dagger} = \sum_{j = 1}^{d^{2}_k} C_k^{ij}\lambda_i^{(k)}
    \label{eq: result}
\end{equation}
The Bloch matrix under the action of $g$ can then be represented as 
\begin{equation*}
    B^{\prime} = (C_1 \otimes C_2) B
\end{equation*}
The characteristic polynomial $f(\Lambda)$ of $B^{\prime}$ can then be expressed as 
\begin{align*}
    f(\Lambda) &= \text{det}((C_1 \otimes C_2)(I_n\Lambda -  B))\\
    &= \text{det}(I_n\Lambda - B)
\end{align*}
This conclusively proves that the charecteristic polynomial and all its coefficients of the Bloch matrix are invariant under determinant-1 SLOCC operations.

\noindent Following from these results we see that the procedure of modification of a Bloch matrix shown in Eq: \eqref{MBM} does not change the set of generators used in the actual Bloch matrix. Hence Eq: \eqref{eq: result} holds true for modified Bloch matrices as well. Thus the eigen values of modified Bloch matrix are invariant under determinant-1 SLOCC operations. 

\vspace{0.2cm}
Now we shall show that the linear homogeneous degree of $MBN$ is less than 4. 

\noindent A function $E()$ that takes pure quantum state $\ket{\psi}$ as input is a linear homogeneous function of degree $\eta$ iff
\begin{equation*}
    E(c\ket{\psi}) = c^{\eta}E(\ket{\psi})
\end{equation*}
where $\eta , c > 0$. 
For a quantum state $\ket{\psi}$ , let its modified Bloch matrix be given as $S_{\psi}$. Then for a quantum state $c\ket{\psi}$ its modified Bloch matrix is represented as, 
\begin{equation*}
    S_{c\psi} = c^2 S_{\psi}
\end{equation*}
This implies that 
\begin{equation*}
    |S_{c\psi}| = c^2 |S_\psi|
\end{equation*}
Since $MBN(\rho)$ has a linear homogeneous degree of $\eta = 2$, it qualifies as an entanglement monotone for pure states.

Any quantum operation on a pure state can be expressed as convex combination of density matrices~\cite{Neilsen_Chuang}. Since $MBN$ is a convex function that is non-increasing under probabilitic LOCC for pure states it directly implies that $MBN$ is non-increasing under probablistic LOCC even for mixed states.

\noindent \textbf{Normalizability:}
For any finite-dimensional bipartite system, a maximally entangled state can be readily expressed in the Schmidt form. This guarantees that $MBN$ is normalizable for all finite dimensions.

Thus $MBN$ satisfies all the essential properties of a valid entanglement measure.

\section{Illustrative examples: Importance of improved discriminance}
$MBN$ is based on a separability criterion that detects certain entangled states that are not detected by standard separability criteria. We demonstrate the importance of this improved discriminance using certain well-known quantum states. (In the following examples, we use  $a= \sqrt{{2}/{M(M-1)}}, \;b = \sqrt{{2}/{N(N-1)}}$, for which  Fei \textit{et al.} \cite{shen2016improved} have shown improvement in  discriminance of the separability criteria (\ref{criteria}). We also fix  $m=4$.) 

\textbf{Example 1: } Conversion of free entanglement into  bound entanglement through mixing is demonstrated in this example. We consider a 2-qutrit state 
\begin{equation}
    \rho({\alpha}) = \frac{2}{7} \ket{\psi_+}\bra{\psi_+} + \frac{\alpha}{7} \sigma_+ + \frac{5-\alpha}{7} \sigma_-,
    \label{horo}
\end{equation}
where
\begin{eqnarray*}
    \ket{\psi_+} & = & \frac{1}{\sqrt{3}}\left( \ket{00} + \ket{11} + \ket{22} \right), \\
    \sigma_+ & = & \frac{1}{3}\left( \ket{01}\bra{01} + \ket{12}\bra{12} + \ket{20}\bra{20} \right), \\
    \sigma_- & = & \frac{1}{3}\left( \ket{10}\bra{10} + \ket{21}\bra{21} + \ket{02}\bra{02} \right).
\end{eqnarray*}
The state parameter \( \alpha \in [2 , 5] \) characterizes the type of entanglement in the state \( \rho({\alpha}) \). The state is separable for \( 2 \leq \alpha \leq 3 \), has bound entanglement for  $3 \leq \alpha \leq 4$ and free entanglement for \( 4 \leq \alpha \leq 5 \). 
Horodecki \textit{et al.}\cite{BE_actv_horo} have shown that the bound entanglement of this state \( \rho({\alpha}) \) can be activated to increase the fidelity of state teleportation. 

\begin{figure}[h!]
    \centering
    \includegraphics[width = \columnwidth ]{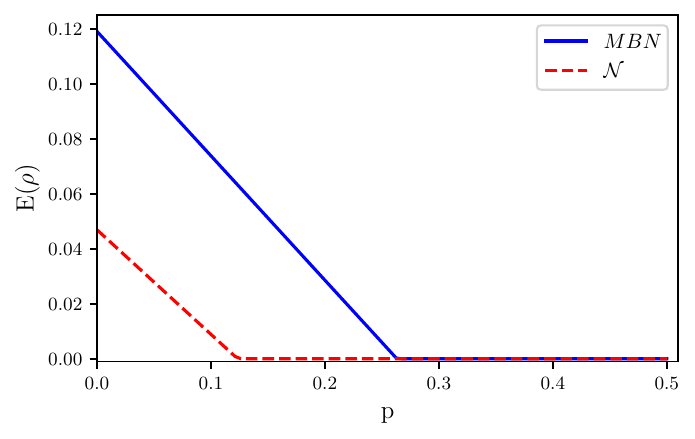}
    \caption{$MBN$ vs $\mathcal{N}$ for $\rho(p,\alpha = 4.5)$ }
    \label{fig-4}
\end{figure}

We mix this state with the maximally mixed state and study the dynamics of entanglement with respect to the mixing parameter $p$: 
\[ \rho(p,\alpha) = p\rho({\alpha}) + (1-p)\frac{\mathbbm{1}}{9}.\]
Fig.~\ref{fig-4} compares  the entanglement dynamics of \( \rho(p,\alpha = 4.5) \), measured by $MBN$ and $\mathcal{N}$.
The phenomenon of  entanglement becoming $0$ abruptly for a  finite value of $p$, observed in Fig: \ref{fig-4}, is known as  Entanglement Sudden Death (ESD).   The entanglement is entirely lost due to environmental noise within a finite time \cite{esd_review}. The ESD time is a critical parameter in the design and implementation of quantum networks, as it determines the time interval within which a quantum state can be utilized effectively in a protocol. Accurate prediction of ESD times ensures the efficient use of quantum states for various tasks. 
From Fig: \ref{fig-4} it is evident that $MBN$ predicts the ESD time of $\rho({\alpha})$ more accurately than $\mathcal{N}$.

\textbf{Example 2:} Next we  consider a particular four qubit state with  bound entanglement, introduced by T\'oth \textit{et al}~\cite{metero_bound}, useful in metrology. We analyse the dynamics of entanglement in this state under  local dephasing. 

Our bipartite state is constructed as a convex combination of six 4-qubit entangled pure states:
(Each 2-qubit subsystem is spanned by the basis states $\{\ket{0} ,\ket{1},\ket{2},\ket{3} \}$) 
\begin{align*}
    \ket{\psi_1} &= \frac{\ket{01} + \ket{23}}{\sqrt{2}}, &\ket{\psi_2} &= \frac{\ket{10} + \ket{32}}{\sqrt{2}}, \\
    \ket{\psi_3} &= \frac{\ket{11} + \ket{22}}{\sqrt{2}}, &\ket{\psi_4} &= \frac{\ket{00} - \ket{33}}{\sqrt{2}}, \\
    \ket{\psi_5} &= \frac{\ket{03} + \ket{12} +\ket{21}}{2\sqrt{2}},& \ket{\psi_6} &= \frac{-\ket{03} + \ket{12}+ \ket{30}}{2\sqrt{2}},
\end{align*}
\begin{equation}
    \rho_{AB} = p\sum_{i=1}^{4} \ket{\psi_i}\bra{\psi_i} + q\sum_{i=5}^{6} \ket{\psi_i}\bra{\psi_i},
    \label{rho_4x4}
\end{equation}
where $q= (\sqrt{2}-1)/2$ and $p = (1-2q)/4$.  

\begin{figure}[h!]
    \centering
    \includegraphics[width = \columnwidth ]{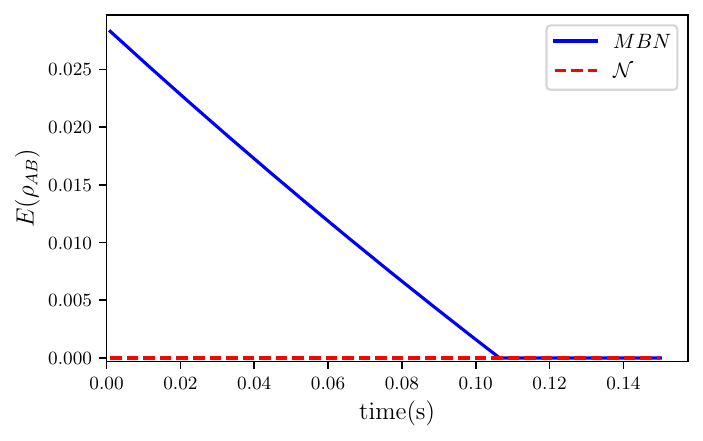}
    \caption{Entanglement dynamics of $\rho_{AB}$ from Eq:~\eqref{rho_4x4} undergoing local phase damping with $\mathcal{T}_2 = 2$ s measured using $MBN$ and $\mathcal{N}$.}
    \label{fig:be_pd}
\end{figure}

Dephasing of this state is modeled using  the  single-qubit Kraus operators
\begin{equation}
K_0(t) = \sqrt{\alpha}\begin{bmatrix}
        1 & 0 \\
        0 & 1
    \end{bmatrix}, \quad K_1(t) =\sqrt{1-\alpha}\begin{bmatrix}
        1 & 0 \\
        0 & -1
    \end{bmatrix}.
\end{equation} 
The dephasing parameter is $\alpha = ({1 + e^{-{t}/{\mathcal{T}_2}}})/{2}$, where $\mathcal{T}_2$ is known as the dephasing time.
The density operator evolves as 
\begin{equation*}
    \rho_{AB}(t) = \sum_{i=1}^{16} \mathcal{K}_i(t)\rho_{AB}(0)\mathcal{K}_i^{\dagger}(t),
\end{equation*}
where the 4-qubit Kraus operators $\mathcal{K}_i$ are the tensor products  
\begin{equation}
  \mathcal{K}_i \in   \left\{\bigotimes_{s = j_1}^{j_{4}} K_s \mid j_1, j_2,j_3,j_{4} \in [0 , 1]\right\}.
\end{equation}
ESD is observed in the time evolution  of \( \rho_{AB} \) as shown in Fig.~\ref{fig:be_pd}, . For this analysis, we have chosen \( \mathcal{T}_2 = 2 \) seconds, a typical value in trapped-ion systems. As observed, the state remains entangled until \( t \approx 0.11 \) seconds while using $MBN$ to measure entanglement while $\mathcal{N}$ is simply zero for all values of $t$.   
This duration $0.11$ seconds is significant for systems like trapped ions, where a typical gate pulse can be applied within \( 20 \, \mu\text{s} \) to \( 100 \, \mu\text{s} \), allowing ample time for entanglement-based operations before decoherence dominates.\\

\textbf{Example 3:}  
Sent{\'{i}}s \textit{et al} have given a Bloch diagonal state $\rho_{BD}$ which would be robust to experimentally prepare and measure~\cite{BE_expt}. We demonstrate that the entanglement  in such a state persists despite subjecting it to a particular form of  noise: correlated amplitude damping.

\begin{figure}[h!]
    \centering
    \includegraphics[width = \columnwidth ]{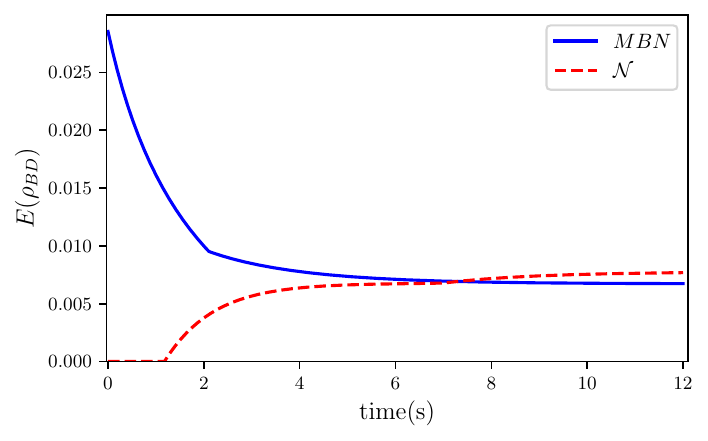}
    \caption{Freezing of entanglement of an initially bound entangled state $\rho_{BD}$ in Eq~\eqref{rho_bd} shown by $MBN$   undergoing correlated amplitude damping with $\gamma = 1$}
    \label{fig:bd_freeze}
\end{figure}

An $N \times N$ bipartite quantum state whose Bloch matrix $B_{AB}$ is diagonal is known as Bloch diagonal state. From Eq:~\eqref{rho_herm}  such a state is represented as 
\begin{equation}
    \rho = \frac{\mathbbm{1}_{N^2}}{N^2} + \sum_{i=1}^{N^2 -1} t_{ii} (\lambda_i \otimes\tilde{\lambda}_i).
\end{equation}
We work with a four qubit state $\rho_{BD}$  with bound entanglement between two qubit bipartitions, whose correlation matrix entries are 
\begin{equation}
    t_{ii} = \left\{ 
    \begin{array}{rl}
    -0.0557066,  &i \in \{1, 2, 3, 4, 5, 8, 11, 13, 15\},  \\
     0.0142664,   &i \in \{6, 10, 12\}, \\
     0.0971467,   &i \in \{7,9,14\}. 
     \end{array} \right.
    \label{rho_bd}
\end{equation}
We study the dynamics of this state using the master equation in the Lindblad form which for pure noise processes is expressed as~\cite{breuer_book}, 
\begin{equation}
   \frac{d\rho}{dt} =  \sum_k \gamma_k\left( L_k \rho L_k^{\dagger} - \frac{1}{2} \{\rho , L_k L_k^{\dagger}\} \right)
\end{equation}
where $L_{k}$ represents the Lindblad operator and $\gamma_k$ is the strength of the noise. 
We consider the correlated amplitude damping channel with one Lindblad operator constructed out of the annhilation operator $\hat{a}$ of qubit energy level:
\begin{equation*}
    L = \hat{a}^{\otimes 4}, \text{ where } \hat{a}= \begin{bmatrix}
   0 & 1 \\
   0 & 0
\end{bmatrix}. 
\end{equation*} 
Fig:~\ref{fig:bd_freeze} shows the dynamics of entanglement of $\rho_{BD}$ when it evolves under correlated amplitude damping channel for $\gamma = 1$. The decay of entanglement of $\rho_{BD}$ reduces dramatically  at $MBN\sim 0.007$ even for longer time scales, exhibiting a  phenomenon which is known as entanglement freezing~\cite{freezing1,freezing2}. 
Eventhough $\mathcal{N}$ is also able to capture the entanglement freezing behaviour, it predicts that the entanglement is zero until $t\approx 1$ seconds implying that the state is useless for various entanglement based operations. 

\noindent Figure~\ref{fig:bd_freeze} also demonstrates that $MBN$ and $\mathcal{N}$ are, in general, non-monotonic with each other under non-local operations. This feature of entanglement measures was predicted and demonstrated for negativity and concurrence by A. Miranowicz \textit{et al.}~\cite{neg_ordering}. This non-monotonicity between $MBN$ and $\mathcal{N}$ clearly indicates that $MBN$ can reveal features of entanglement that negativity could never capture.


Apart from the above examples, we have verified that $MBN$ is non-zero for various  families of bound entangled states in the literature \cite{be1, be2, be3, be4,BE_dist}.   Thus $MBN$ serves as an efficient tool to study dynamics of bound entanglement under various conditions potentially enabling the observation of various exotic behaviors of bound entanglement such as entanglement freezing, rebirth\cite{rebirth}, thawing\cite{freezing2} and so on. 

\begin{figure*}[t]
    \centering
    \includegraphics[width=8cm]{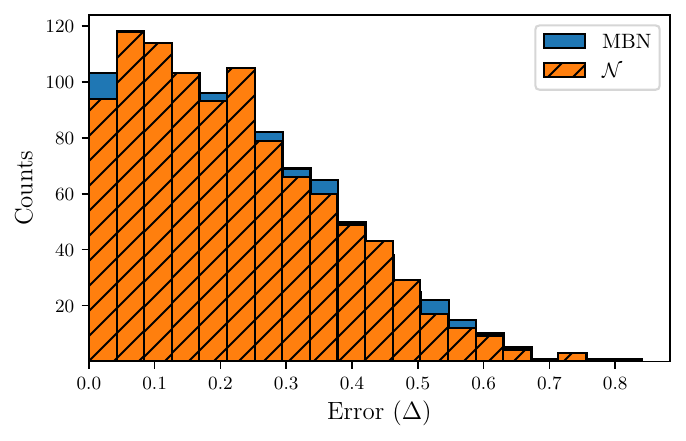}
    \caption*{(a) Two qubit ME states (n = 100)}
\end{figure*}
\begin{figure*}[t]
    \begin{tabular}{c c}
    \includegraphics[width=8cm]{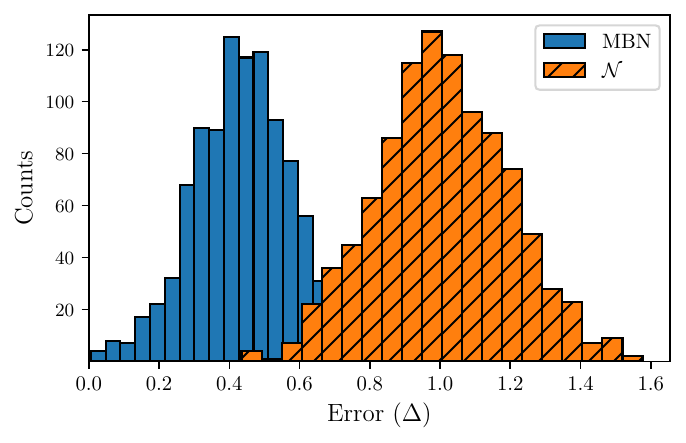} & \includegraphics[width=8cm]{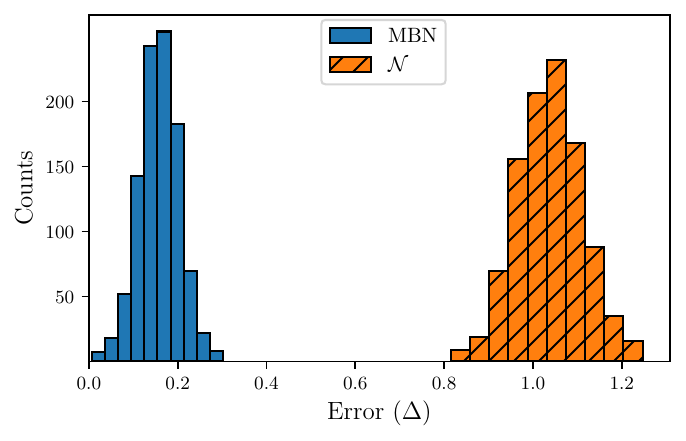} \\
    (b) Three qubit ME states (n = 700) & (c) Four qubit ME states (n = $2\times 10^4$)  \\ 
     & \\ 
    \end{tabular}
\caption{Frequency distribution of error (\(\Delta\)) for Negativity ($\mathcal{N}$) and $MBN$ using \( n \) copies of 1,000 LU-equivalent states of (a) \( \ket{\psi_2} \), (b) \( \ket{\psi_3} \), and (c) \( \ket{\psi_4} \) in Eq:\eqref{ames} . }
\label{error_by_tomography}
\end{figure*}
 \section{Robustness under inaccurate estimation of quantum state}
 In the introduction, we discussed the need for and various approaches to estimate entanglement using finite copies of quantum states.

We would like to ask an important question in this regard:  
\textit{Do all entanglement measures suffer equally from inaccuracies in state estimation using Linear Inversion(LI)?}

In attempting to answer this question, it is important to note that no analytical solution exists. The probabilistic nature of measurement outcomes makes the associated errors inherently random. Thus, the only meaningful approach is to generate a large sample of random quantum states and address the question quantitatively.

The notion of Maximal Entanglement (ME) for multi-qubit systems has been defined by G. Gour \textit{et al}~\cite{Gour_max_ent}.
Such states for two qubits, three qubits, and four qubits in the computational basis are expressed as:
\begin{align}
    \ket{\psi_2} &= \frac{1}{\sqrt{2}}\ket{00} + \ket{11}, \nonumber \\
    \ket{\psi_3} &= \frac{1}{\sqrt{2}} \left( \ket{000} + \ket{111} \right),  \label{ames} \\ 
    \ket{\psi_4} &= \frac{1}{2}\left( \ket{0000} + \ket{1100} + \ket{0011} - \ket{1111} \right) \notag.
\end{align}
We calculate entanglement between the first qubit and the rest of the system using negativity ($\mathcal{N}$) and $MBN$ which we  denote as $E_{\text{true}}$.
We then simulate state estimation through LI using n copies of the state under consideration to obtain an estimated Hermitian matrix $\rho_{\text{est}}$. We  calculate the entanglement of $\rho_{\text{est}}$ which is represented as $E_{\text{expt}}$.
To ensure that our results were not biased by the choice of any particular state, we generated 1,000 LU-equivalent states for each of the above states. The estimated entanglement from this process is \( E_{\text{expt}} \), and the  error in entanglement quantification is measured as
\begin{equation*}
    \Delta = \frac{|E_{\text{true}} - E_{\text{expt}}|}{E_{\text{true}}}.
\end{equation*}

In Fig.~\ref{error_by_tomography}, we present the frequency distribution of error \( \Delta \). 
For the two-qubit case, the frequency distribution of error is nearly identical for $MBN$ and $\mathcal{N}$ when 100 copies of the state are used for characterization. However, for three- and four-qubit systems, the frequency distribution of \( \Delta \) for $MBN$ is centered around significantly lower values compared to $\mathcal{N}$. This clearly indicates that, when finite copies of a quantum state are used to quantify entanglement via Linear Inversion (LI), $MBN$ exhibits lower error rates than $\mathcal{N}$, with this trend becoming more pronounced as the number of qubits increases.

These results are justified by the following.
As mentioned in the introduction, linear inversion  is a straightforward technique for estimating quantum states. However, it may yield Hermitian matrices with negative eigenvalues, that are therefore  not valid density operators. 

Since negativity quantifies entanglement based on the negative eigenvalues of partially transposed density operators, the inaccuracies in the LI-estimated density matrix propagate to the computed negativity, often resulting in significant errors. 

In contrast, $MBN$ relies on the deviation from a norm value characteristic of separable states. This measure is inherently more robust, 
thereby reducing the impact of errors on the entanglement estimate.

The difference in the frequency distribution of errors  in $\mathcal{N}$ and $MBN$ naturally increases with system size, as the number of negative eigenvalues will also increase with system size ~\cite{Avron_2019}.

\section{Conclusion}
 In this work, we introduce a bipartite entanglement measure, the Modified Bloch Norm ($MBN$) derived from the Improved Bloch Matrix (IBM) separability criterion. We show that this measure satisfies the desirable properties for effective entanglement quantification. Through practical examples, we illustrate how $MBN$ can estimate the Entanglement Sudden Death (ESD) time with greater accuracy than the widely used entanglement measure   Negativity $\mathcal{N}$. This improvement could enable the study of bound entanglement dynamics in complex environments, such as those involving non-Markovian noise, potentially revealing unobserved aspects of bound entanglement.

We further demonstrate that $MBN$ achieves significantly lower error rates than Negativity when the quantum state is estimated using Linear Inversion (LI). For experimental scenarios where accurate entanglement estimation is needed and Maximum Likelihood Estimation (MLE) or Bayesian Inference (BI) are  computationally intensive, $MBN$ offers a more efficient alternative for estimating entanglement reliably.
 
\noindent \textbf{Author Contributions:} All authors contributed equally to this work. The authors declare that there are no conflicting ideas among them regarding the content of this manuscript.

\section{Acknowledgements}
\noindent We would like to acknowledge support from the Department of Science and Technology, Government of India, through the project DST/ICPS/QuST/Theme-3/2019/Q109.\\
We  thank Askar Ali for his valuable discussions in linear algebra.
All calculations and simulations in this work were performed using the QuTiP library \cite{qutip1,qutip2}.

\bibliography{GeometryReferences.bib}

\end{document}